# Reducing the complexity of computing the values of a Nash equilibrium.

**Debtoru Chatterjee[1] • Girish Tiwari[2] • Niladri Chatterjee[3]**

## Abstract

The Colonel Blotto game, formulated by Émile Borel, involves players allocating limited resources to multiple "battlefields" simultaneously, with the winner being the one who allocates more resources to each battlefield. Computation of the Nash equilibrium, including of two-person, zero-sum, mixed strategy Colonel Blotto games have encountered issues of scalability and complexity owing to their PPAD completeness. This paper proposes an algorithm that computes the same value as the Nash equilibrium but cannot be characterized by the Fixed-Point Theorems of Tarski, Kakutani and Brouwer. The reduced complexity of the proposed algorithm is based on dispensing with the need for computing both players' Nash strategies in Colonel Blotto games. The same algorithm can, therefore, be extended to all two-person, zero-sum games to compute the value of the Nash equilibrium. The theoretical superiority of the proposed algorithm over both LP solvers and another method that computes the same value of the game as its Nash equilibrium by a random assignment of probabilities to the active strategy set of the defending player, is also proposed.

## 1.      Introduction

A Colonel Blotto game is a two-person zero-sum game in which the players are required to simultaneously distribute limited resources over more than one object. The player allocating the greater resource to the object wins it. The game was formulated by the French mathematician Emile Borel in 1921. It was published by Borel and Jean Ville as an application of the Theory of Probability and of the Game of Chance in 1938. Popularized by Operations Research scholars after the Second World War, Borel's game became and remains a classic in the literature of Game Theory under the name of the Colonel Blotto Game.

Strategic resource allocation using the Colonel Blotto game has applications not only in military settings, but also in politics, network security, hiring decisions and even competition for social media attention. Its pre-eminent use, however, has been in the field of security and conflict, involving allocation of defense resources to a certain number of 'battlefields' that is the object of contest between adversaries.

Debtoru Chatterjee, debtoruchatterjee@gmail.com, a Joint Secretary to the Government of India, is a doctoral candidate at the Indian Institute of Technology, School of Interdisciplinary Research, New Delhi, 110016. He conceived all notions expressed in the paper, their analysis and presentation, and wrote the text.

Girish Tiwari assisted the first author in linear programming with nashpy and in editing.

Niladri Chatterjee, niladri.chatterjee@maths.iitd.ac.in, is a Professor of Computer Science and Statistics in the Department of Mathematics, Indian Institute of Technology (IITD), New Delhi, 110016, India. He examined and edited the paper.

## 2. Literature review and motivation

Chen, Deng and Teng (2009), Goldberg Papadimitriou and Savani (2011) showed that computing a Nash equilibrium in a Colonel Blotto game was PPAD complete owing to the CB game's structural properties of non-convexity and non-linearity. Daskalakis, Goldberg, Papadimitriou (2008) proved the reduction of the Brouwer Fixed Point Theorem to the polynomial parity argument for directed graphs (PPAD) class and then showed the reduction of Nash to Brouwer to prove that Nash was in PPAD. Nash in his 1950 proof had himself relied on both the Brouwer and Kakutani Fixed Point Theorems to show that a mixed strategy Nash Equilibrium existed for finite games.

We show, however, in this paper, that, using our proposed algorithm, it is possible to compute the Nash Equilibrium value of a two person, zero sum Colonel Blotto Game even while deviating from the framework and requirements of the Kakutani Fixed Point Theorem and, by extension of discrete strategies spaces to continuous strategies spaces, also while deviating from the framework and requirements of the Brouwer Fixed Point Theorem.

Roberto Ricci used Tarski's Fixed-Point Theorem to prove the existence of pure strategy Nash Equilibria for two player games with possible discontinuous payoff functions defined on compact real intervals.

We prove that the Nash Equilibrium value of the two persons, zero sum Colonel Blotto Game can be computed even by deviating from the framework of the Tarski Fixed Point Theorem.

Ahmadinejad et al. (2016) discussed the scalability issues in computing Nash equilibria for large Colonel Blotto games. They proposed a heuristic approach to reduce the game's computational complexity. The authors suggested: (a) sampling a smaller sub-set of strategies rather than considering the entire strategy space; (b) applying k means clustering to group similar strategies together; (c) computing a representative mixed strategy for each cluster; and (d) using these representative strategies to compute the Nash equilibrium.

By adopting this heuristic approach, Ahmadinejad et al. demonstrated improved computational efficiency in computing Nash equilibria for large Colonel Blotto games. However, they acknowledged the potential issue of dominant strategies being thus sampled out. Yet they did not address this serious concern in their proposed heuristic approach.

To mitigate the issue of dominant strategies being sampled out thus affecting the accuracy of the approximated Nash equilibrium, the authors could have considered techniques like stratified sampling from different regions of the strategy space to ensure dominating strategies were not overlooked. They could have assigned higher sampling probabilities to strategies that were more likely to be part of the Nash equilibrium. Alternatively, they could have analyzed the sampled strategies to identify and reintroduce the dominating strategies that might have been sampled out.

Such mitigation approaches would be, however, computationally much more complex and intensive than the algorithm proposed in this paper to compute, without loss of accuracy, the Nash equilibria for large Colonel Blotto games more efficiently than any other algorithm till date.

Nashpy, which implements the Lemke-Howson algorithm, computes the Nash equilibrium strategies for both players, even though the game's value remains unaffected if only the row player adheres to his Nash equilibrium strategy while the column player (defender) chooses each of his active strategies in his strategy set with any random probability (provided the sum of such probabilities is 1). This is a matter of convention and tradition as well as to obviate randomization errors stemming from leaving the column player to randomly decide the probability of playing each of his active strategies.

Computing both players' Nash equilibrium strategies is arguably redundant: that this approach provides a more comprehensive theoretical understanding of the CB game or any two-person zero sum game is debatable. On the other hand, as demonstrated in this paper, such a conventional approach both obviously and provably adds to the complexity of calculating the Nash equilibria values. The Lemke-Howson algorithm, which is used to calculate the Nash equilibrium value of the game (including in Colonel Blotto games) is not strongly polynomial (Kristina Kask, Paul W. Goldberg, and Bernhard von Stengel, 2011) and neither is its counterexample (Bernhard von Stengel, 1999). The linear programming approach to finding the Nash Equilibrium of two-person, zero sum games involves algebraic optimization over the strategies of both row and column and becomes increasingly complex over large game trees. Papers by researchers like Sandholm (2015) have explored various methods to reduce computational complexity, including linear programming relaxations. However, no one has specifically highlighted the redundancy of computing both players' Nash Equilibrium strategies in Colonel Blotto games.

This paper introduces an algorithm that reduces the computational complexity involved in the computing of the values of the game without introducing randomization errors or excluding dominant strategies. Its superiority over other approaches is also demonstrated. We have already mentioned above that the use of our algorithm results in a computation of the Nash Equilibrium values of not only the Colonel Blotto game but also of all two-person, zero sum games beyond the framework or requirements of the Brouwer, Kakutani and Tarski Fixed Point Theorems.

## 3.    The Algorithm

According to the solution theory proposed by Nash, an equilibrium was an action profile $a^*$ with the property that no player $i$ could do better by choosing an action different from $a_i^*$, given that every other player $j$ adhered to $a_j^*$. Such an action profile, $a^*$, required *each* player to play each of his active strategies with a probability determined by the Nash equilibrium. In precise terms, the action profile $a^*$ in a strategic game with ordinal preference was a **Nash equilibrium** if, for every player $i$ and every action $a_i$ of player $i$, $a^*$ was at least as good according to player $i's$ preferences as the action profile $(a_i, a^*_{-i})$ in which player $i$ chose $a_i$ while every other player $j$ chose $a_j^*$. Equivalently, for every player $i$,

$$u_i\ (a^*) \geq\ u_i\ (a_i, a^*_{-i})\ for\ every\ action\ a_i\ of\ player\ i \tag{1}$$

Given this Nash equilibrium notion in two-person, zero sum games, Colonel Blotto mixed strategy games, which are a class of two-person, zero sum games have a Nash equilibrium conforming to the above.

The proposed algorithm is introduced through the following proposition:

**Proposition 1**: In a set of Colonel Blotto games $B$ (and all other two-person, zero sum games), where the resources $r$ of row player $R$ are in the range of $1 \leq r \leq$ x, the resources $c$ of the column player $C$ are in the range of $1 \leq c \leq$ y and the row player $R$ (attacker) adheres to his action $a_j^*$ in the action profile $a^*$:

$$u_i\ (a^*) = u_i\ (a_i, a^*_{-i})\ for\ every\ action\ a_i\ of\ player\ i \tag{2}$$

where,

Column player $C$ (defender) represent player $i$; $a_i$ is a set of active strategies $n$ in a mixed strategy game
such that each element of the set of active strategies $a_i = \{x_1, x_2 \ldots x_n\}$ is played with the probability $1/n$; and

$a^*$ is the action profile of each player given by the Nash equilibrium wherein the active strategies of each player are spread over a probability distribution *determined by the Nash equilibrium*.

Thus, while Nash equilibrium theory requires *each* player to play each of his active strategies *according to the probability determined by the Nash equilibrium,* our algorithm proposes that even when only the row player plays each of his active strategies according to the probability determined by the Nash equilibrium while the column player *deviates from this rule by playing each of his active strategies with an **equal** probability*, the **same value** of the game as the Nash equilibrium is obtained. Thus, under the Nash equilibrium theory, while the probability distributions of row and column's active strategies may or may not be uniform since they are Nash determined, as per our proposed algorithm, the probability distribution of column's active strategies is required to be ***always uniform*** (with no reference to Nash) but in association with the Nash equilibrium determined probability distribution of row. Thus, the proposed algorithm incorporates Nash only halfway in respect of row but not in respect of column.

It follows logically that our proposed algorithm reduces the complexity of computing the values of the Nash equilibrium by eliminating the need to compute the Nash equilibrium–determined probability distributions of *both* row and column. Instead only the Nash equilibrium-determined probability distribution of row is computed, while the column's active strategies (n) are played according to a simple, uniform probability of $1/n$ for each active strategy. This *saving of compute power yields the same value of the game as the Nash equilibrium*.

The mathematical proof of the forgoing proposition is given in the **Appendix 'A'**.

Secondly, it is also seen that in a set of Colonel Blotto games $B$ and all other two-person, zero sum games, where the resources $r$ of row player $R$ are in the range of $1 \leq r \leq x$, the resources $c$ of the column player $C$ are in the range of $1 \leq c \leq y$ and the row player $R$ adheres to his action $a_j^*$ in the action profile $a^*$:

$$u_i\ (a^*) = u_i\ (a_i, a^*_{-i})\ \textit{for every action } a_i \textit{ of player } i \tag{2}$$

where,

Column player $C$ represent player $i$; $a_i$ is a set of active strategies $n$ in a mixed strategy game such that each element of the set of active strategies $a_i = \{x_1, x_2 \ldots x_n\}$ is played with a probability such that $\sum_{i=1}^{n} x_n = 1$; and

$a^*$ is the action profile of each player given by the Nash equilibrium wherein the active strategies of each player are spread over a probability distribution *determined by the Nash equilibrium*.

What the above implies is that while Nash equilibrium theory requires *each* player to play each of his active strategies *according to the probability determined by the Nash equilibrium,* the above method proposes that even when only the row player plays each of his active strategies according to the probability determined by the Nash equilibrium while the column player *deviates from this rule by playing each of his active strategies with any probability summing to 1*, the **same value** of the game as the Nash equilibrium is obtained.

Thus, this method is in competition with our proposed algorithm in computing the same value as the Nash equilibrium without the need for computing the Nash equilibrium-determined probability distributions of *both* row and column. But while our proposed algorithm requires column's active strategies (n) to be played according to a simple, uniform probability of 1/n for each active strategy, the above method requires column's active strategies to be played with any probability for each strategy, summing to 1 for all its active strategies. Both our proposed algorithm and the above method retain the requirement for row's active strategies to be played over a Nash equilibrium-determined probability distribution. Thus, both incorporate Nash in respect of row but not in respect of column.

This paper proposes to show that the complexity of computing the value of the Nash equilibrium according to our algorithm is less than the complexity of computing the value of the Nash equilibrium according to above method.

Lastly, optimization through linear programming (such as the LP solver in pyspiel) is another computationally efficient algorithm that computes the same value as the Nash equilibrium. It is polynomially far more efficient than the Lemke-Howson algorithm. However, its assignment of probabilities to the active strategies of row and column is based on algebraic optimization methods. LP solvers, including pyspiel, cannot be characterized by fixed point theorems such as those of Brouwer, Kakutani and Tarski, which are known to be in PPAD. I propose that our algorithm too deviates from the framework of the above fixed point theorems. However, I show that the computational complexity of the proposed algorithm is less than that of the Linear Programming solver of two-person, zero sum games, including Colonel Blotto games. I also propose other advantages of our algorithm over an LP solver in terms of maximum entropy, robustness to mistakes (Selten's 'Trembling Hands' theorem), robustness to incomplete information (Aumann), and Minimax Regret characterization (Rubenstein and Osborne). The superiority of our approach (equal probabilities over active column strategies) over the 'probabilities summing to 1 over active column strategies' approach on all the above counts is also established.

The paper is organised as follows. Section 4 gives a full description of the proposed algorithm, its functioning and the intuition behind it. Section 5 proves the computational efficiency of our algorithm vis-a-vis Linear Programming solvers, the Lemke Howson algorithm (which computes the Nash Equilibrium using the tradition method of determining the NE of both row and column) and the 'Probabilities summing to 1' method. Sections 6,7,8 & 9 propose the superiority of our algorithm over both LP solvers and the 'Probabilities summing to 1' method on the scores of maximum entropy, robustness to mistakes, minimax regret characterization and incompleteness of information. Section 10 summarizes the discussion and Section 11 suggests the lines of further research. Appendix 'A' gives the mathematical proof of proposition 1 (equal probabilities over active column strategies). Appendix 'B'gives a brief overview of how the proposed algorithm deviates from the framework of the Brouwer, Kakutani and Tarski fixed point theorems.

## 4.    Functioning of the proposed algorithm

The following examples are taken from the Colonel Blotto games given by Philip D. Straffin (Straffin 1993) under the name of 'Guerrillas vs. Police'. A model is constructed consisting of *g* guerrillas, *p* policemen and two government arsenals which the guerrillas seek to capture and the police needs to defend. The guerrillas win if they capture any one of the two arsenals. (They can thus increase their arms to sustain their conflict with government forces). The police, however, win only if they are able to successfully defend both arsenals.

The guerrillas win if $g > p$. The police win when $p \geq 2g$ as they need to defend each of the two arsenals with a force of at least $g$. The outcomes when $g \leq p < 2g$ are examined below.

Assuming $g = 4$ and $p = 4$, Straffin depicts the following payoffs to the guerrillas (1 for a win 0 for a loss):

**Table 1** Symmetric mixed strategy Colonel Blotto game

|  |  | 4 Police | | |
|---|---|---|---|---|
|  |  | 4-0 | 3-1 | 2-2 |
| 4 Guerrillas | 4-0 | ½ | 1 | 1 |
|  | 3-1 | 1 | ½ | 1 |
|  | 2-2 | 1 | 1 | 0 |

The guerrillas can opt to divide 4-0, 3-1, 2-2, its three active strategies. Police can similarly divide 4-0, 3-1 and 2-2 to protect the arsenal(s). The payoffs to the guerrillas are calculated as follows:

Case 1: Guerrillas divide 4-0 and police divide 4-0. The expected payoff is ½, since the guerrillas attack the defended arsenal with probability 0.5 and lose. But also they attack the undefended arsenal with a probability of 0.5 and win.

Case 2: Guerrillas divide 4-0 and police divide 3-1. Here the expected payoff for guerillas is 1. They win both the arsenals each of which they attack with a probability of ½.

Case 3: Guerrillas divide 4-0 and police divide 2-2. The guerrillas get a payoff of 1. They win both the arsenals each of which they attack with a probability of ½.

Case 4: Guerrillas divide 3-1 and the police divide 4-0. The guerrillas get a payoff of 1. They win when they attack, with probability of ½, the undefended arsenal with a party of 3 guerrillas and when they attack, with a probability of ½, the undefended arsenal with a party of 1 guerrilla.

Case 5: Guerrillas divide 3-1 and the police divide 3-1. The guerrillas get a payoff of ½. They win when the party of 3 guerillas attacks the arsenal defended by 1 policeman with a probability of ½.

Case 6: Guerrillas divide 3-1 and police divide 2-2. The guerrillas get a payoff of 1. They win when the party of 3 guerillas attacks each arsenal with a probability of ½.

Case 7: Guerrillas divide 2-2 and police divide 4-0. The guerrillas get a payoff of 1. Each party of 2 guerrillas wins by attacking the undefended arsenal with a probability of ½.

Case 8: Guerrillas divide 2-2 and police divide 3-1. The guerrillas get a payoff of 1. Each party of 2 guerrillas wins by attacking the arsenal defended by 1 policeman.

Case 9: Guerrillas divide 2-2 and police divide 2-2. The guerrillas get a payoff of 0. Each party of 2 guerrillas is defeated by 2 policemen at each arsenal.

The value of the game is calculated by multiplying each payoff by the Nash equilibrium-determined probability with which the guerrillas and police will play their relevant row and column strategy respectively and then summing the products. In this Colonel Blotto game, the value of the game indicates the probability with which the guerrillas would win. For games involving multiple rows and columns, *Nashpy*, Release 0.0.41, a python library for 2 player games, was used to calculate the probabilities with which *both* players will play their different strategies as well as the value of the game. The Nash equilibrium-determined probabilities with which each player will play each of his or her strategies and the value of the game are given in the program below:

**Figure 1**

```
1  # 4 Guerillas and 4 Policemen
2  import nashpy as nash
3  import numpy as np
4
5  # Define the game
6  A = np.array([[0.5, 1, 1],
7                [1, 0.5, 1],
8                [1, 1, 0]
9                ])
10 rps = nash.Game(A)
11
12 # Find Nash equilibria
13 eqs = rps.support_enumeration()
14
15 # Extract and print the first equilibrium
16 row_player_strategy, column_player_strategy = next(eqs)
17 print("Row Player Strategy:", row_player_strategy)
18 print("Column Player Strategy:", column_player_strategy)
19
20 # Calculate the value of the game
21 value_of_game = row_player_strategy.dot(A).dot(column_player_strategy)
22
23 print("Value of the Game:", value_of_game)
```

```
Row Player Strategy: [0.4 0.4 0.2]
Column Player Strategy: [0.4 0.4 0.2]
Value of the Game: 0.8000000000000002
```

Nash equilibrium computes a probability of 0.4, 0.4 and 0.2 to each of the guerrillas' 3 strategies. *It also computes a probability of 0.4, 0.4 and 0.2 to each of the 3 police strategies* and computes the value of the game as 0.80.

As per the proposition, the police (the defender) ought to play each of its *active* strategies with a probability of 1/n, where n is the number of active strategies. Hence, in the above game, Police ought to play each of its 3 active strategies with a probability of 0.33 (rounded of to 0.34 for any one of the three strategies. The number of decimal places may be extended without affecting the result computed to any significant decimal place). Using this proposition, the value of the game is recalculated as follows, retaining the payoffs to the guerrillas and the Nash equilibrium-determined probabilities with which the latter (guerrillas) plays each of its strategies:

**Table 2**

| | | |
|---|---|---|
| 0.5 x 0.4 x 0.33 | = | 0.066 |
| 1 x 0.4 x 0.33 | = | 0.132 |
| 1 x 0.4 x 0.34 | = | 0.136 |
| 1 x 0.4 x 0.33 | = | 0.132 |
| 0.5 x 0.4 x 0.33 | = | 0.066 |
| 1 x 0.4 x 0.34 | = | 0.136 |
| 1 x 0.2 x 0.33 | = | 0.066 |
| 1 x 0.2 x 0.33 | = | 0.066 |
| **Total** | **=** | **0.8** |

*It is thus seen that our proposed algorithm dispenses with the requirement of the Police to separately compute the Nash-equilibrium probabilities of its strategies and yet achieves the same value of the game because the Guerilla adheres to his Nash-equilibrium strategy.*

One philosophy behind this algorithm is that when the above Colonel Blotto game is played according to the proposed algorithm, the defender (Police) inflicts greater uncertainty on the attacker (guerilla) than is achieved when the former plays his active strategies according to the Nash equilibrium-based probabilities. We prove this statement later in this paper.

The intuition behind column's (defender's) above approach is that when row plays Nash Equilibrium (NE) probabilities to make column indifferent between column strategies, column **neutralizes** this by playing column's active strategies with equal probability. Column **does not** try to manipulate by assigning probabilities so that row becomes indifferent between row strategies. Hence, column **does not** play NE probabilities because he does not reciprocate row's manipulation in the same coin. There is no need to do so. He simply neutralizes row by playing each of his (column) active strategies with an equal probability.

We now demonstrate that the proposed algorithm is robust to a larger Colonel Blotto game involving dominating and dominated strategies Assuming $g = 8$ and $p = 9$, the payoffs to the guerrillas are depicted in the following table:

**Table 3** Asymmetric mixed strategy Colonel Blotto game

|              |     | 9 Police |         |         |         |         |
|--------------|-----|----------|---------|---------|---------|---------|
|              |     | 9-0      | 8-1     | 7-2     | 6-3     | 5-4     |
|              | 8-0 | ½        | ½       | 1       | 1       | 1       |
|              | 7-1 | 1        | ½       | ½       | 1       | 1       |
| 8 Guerrillas | 6-2 | 1        | 1       | ½       | ½       | 1       |
|              | 5-3 | 1        | 1       | 1       | ½       | ½       |
|              | 4-4 | 1        | 1       | 1       | 1       | 0       |

Inputting the payoffs to the guerrilla in the payoffs matrix of the *Nashpy* program, the probabilities attached to the different strategies of the guerrillas and police and the value of the game are computed as follows:

**Figure 2**

```python
# 8 Guerillas and 9 Policemen
import nashpy as nash
import numpy as np

# Define the game
A = np.array([[0.5, 0.5, 1, 1, 1],
              [1, 0.5, 0.5, 1, 1],
              [1, 1, 0.5, 0.5, 1],
              [1, 1, 1, 0.5, 0.5],
              [1, 1, 1, 1, 0]])
rps = nash.Game(A)

# Find Nash equilibria
eqs = rps.support_enumeration()

# Extract and print the first equilibrium
row_player_strategy, column_player_strategy = next(eqs)
print("Row Player Strategy:", row_player_strategy)
print("Column Player Strategy:", column_player_strategy)

# Calculate the value of the game
value_of_game = row_player_strategy.dot(A).dot(column_player_strategy)

print("Value of the Game:", value_of_game)
```

```
Row Player Strategy: [0.4 0.  0.4 0.  0.2]
Column Player Strategy: [0.  0.4 0.  0.4 0.2]
Value of the Game: 0.8000000000000002
```

After crossing out dominated strategies, it is seen that the police are left with 3 *active* strategies and the guerrillas are also left with 3 active strategies. The resultant payoff matrix exactly represents the payoff matrix depicted in Table 1. Applying the probability of 1/n to each of n *active* strategies of police and retaining the Nash equilibrium-determined probabilities attached to the strategies of the guerrillas, the same value of the game of 0.80 is achieved.

The table below is a small sample of Colonel Blotto games, both symmetric and asymmetric, and including games with and without dominating strategies. The value of the game achieved by the Nash solution and the proposed algorithm are the same in each case, demonstrating the robustness of the algorithm:

### Value of Colonel Blotto game

|  | Proposed algorithm | Nash solution |
|---|---|---|
| 9 Policemen and 8 Guerillas | 0.80 | 0.80 |
| 5 Policemen and 5 Guerillas | 0.83 | 0.83 |
| 4 Policemen and 4 Guerillas | 0.80 | 0.80 |
| 6 Policemen and 5 Guerillas | 0.66 | 0.66 |
| 8 Policemen and 8 Guerillas | 0.88 | 0.88 |
| 11 Policemen and 10 Guerillas | 0.83 | 0.83 |
| 7 Policemen and 6 Guerillas | 0.75 | 0.75 |
| 7 Policemen and 7 Guerillas | 0.875 | 0.875 |
| 6 Policemen and 6 Guerillas | 0.857 | 0.857 |
| 9 Policemen and 9 Guerillas | 0.90 | 0.90 |
| 10 Policemen and 9 Guerillas | 0.80 | 0.80 |

## 5.    Computational efficiency and simplicity

Computational efficiency and computational simplicity in game theoretic algorithms have been discussed by C.H. Papadimitriou (2001) To demonstrate the superiority of the proposed algorithm, we show that it is superior to an algorithm where the defender (column) plays each of his active strategies with any random probability so long as the sum of such probabilities is 1. This comparison is made because when the column player plays each of his active strategies with any random probability summing to 1 (instead of with equal probability as in our proposed algorithm) the same value as the Nash equilibrium is computed (as by our proposed algorithm). We name this algorithm of random probabilities summing to 1 as a Probabilities Sum Approach (PSA). Our own algorithm is abbreviated as Equal Probabilities Approach (EPA). First we demonstrate that PSA is computationally more efficient than the algorithm which computes the defender's Nash Equilibrium strategy, given that the attacker's (row's) Nash Equilibrium strategies have to be calculated in both the EPA approach and the PSA approach.

**Proposition 2:** *The computational complexity of calculating the Column Nash Equilibrium Strategy in a Colonel Blotto (or any two person, zero sum) game by the Lemke-Howson algorithm (or by any method that calculates the Nash Equilibrium strategy of Column) is more than or at best equal to the PSA approach.*

**Proof:**

Colonel Blotto Game:

- n row strategies (i = 1, 2, ..., n)
- m column strategies (j = 1, 2, ..., m)
- Payoff matrix: $A = [a_{ij}]$, where $a_{ij}$ is the payoff to Row when playing i against Column's j

The computational complexity of the Lemke-Howson algorithm for calculating the column Nash equilibrium is *$O(2^m)$*, exponential in column strategies, in the worst-case scenario according to Savani R. & von Stengel.B (2004) and, using heuristics, at best *$O(m)$,* linear in all column strategies and not just active ones, according to Codenotti B et al. (2008).

**PSA Approach:**

In the PSA approach, Column plays each active strategy with a random probability, subject to the constraint that the probabilities sum to 1:

$$\sum_{j=1}^{m} q_j = 1$$

$$q_j \geq 0, \forall j$$

Where $q_j$ is the probability of column player's strategy. The computational complexity of generating random probabilities that satisfy these constraints is $O(m)$.

**Comparison:**

Since the computational complexity of calculating the Column Nash Equilibrium Strategy is between $O(m)$ and $O(2^m)$ while the computational complexity of the PSA approach is $O(m)$, we can conclude that:

The computational complexity of calculating the Column Nash Equilibrium Strategy by the Lemke-Howson algorithm is at best the same as that of the PSA approach.

In other words, the PSA approach is computationally at least as efficient as the Lemke-Howson algorithm (or any method that computes column's NE strategies) for calculating the Column Nash Equilibrium Strategy. It needs to be reiterated that the least complexity in computing the Nash-equilibrium strategy for the column player, $O(m)$, is attained by the Lemke-Howson algorithm only by the adoption of heuristics. On the other hand, the PSA approach attains the same level of computational complexity, $O(m)$, without recourse to heuristics.

We now prove that the EPA approach requires fewer calculations than the PSA approach making it a more computationally efficient strategy.

**Proposition 3:** *The computational complexity of the PSA approach as well as of the linear programming approach is more than that of the EPA approach.*

**Proof:**

As shown earlier, in the PSA approach, Column plays each active strategy with a random probability, subject to the constraint that the probabilities sum to 1:

$$\sum_{j=1}^{m} q_j = 1$$

$$q_j \geq 0, \forall j$$

Where $q_j$ is the probability of the column player's strategy. The computational complexity of generating random probabilities that satisfy these constraints is $O(m)$.

Let's denote the EPA approach as $\sigma_{EPA}$, where $\sigma_{EPA,i} = 1/n$ for all $i$ column strategies.

The mathematical complexity of the EPA approach can be measured by the number of calculations required to compute $\sigma_{EPA,i} = 1/n$. The complexity is thus $O(1)$. Column does not solve a best-response LP or compute a Nash Equilibrium

In general, the PSA approach requires more calculations than the EPA approach, *especially when the number of strategies is large.*

Hence it also stands proved that EPA is computationally more efficient than the Lemke-Howson method of computing the Nash Equilibrium strategies of both the defender and the attacker in Colonel Blotto Games and in all two-person, zero sum games.

Linear programming based complexity of computing the optimal column strategy in a two-person, zero sum game has a bound that is $\boldsymbol{O((n+m)^{3.5} \cdot \log L)}$, according to Karmarkar, where,

**m** represents the number of rows (strategies for the row player i.e. attacker),

**n** represents the number of columns (strategies for the column player i.e. defender),

**L** represents the number of bits required to represent the input data.

This complexity formula thus indicates that as **m** and **n** increase (more rows and columns due to larger resource units of row and column player), the complexity grows polynomially. However, as **L** increases (more bits required to represents the input data), ***the complexity grows exponentially with respect to L.***

Thus for large, two person zero sum games, including large Colonel Blotto games, the complexity of computing the optimal column (defender) strategy, using Karmarkar algorithm, indeed increases significantly both due to the polynomial growth with respect to **m** and **n and** the ***exponential*** growth with respect to **L**.

The complexity of eliminating dominated strategies from the payoff matrix of two person zero sum games, including Colonel Blotto games, required by our Equal Probabilities algorithm however, grows ***quadratically*** with respect to the number of rows and columns, but since the number of comparisons depends on both **m** and **n,** ***the overall complexity grows polynomially, specifically O(m\* n\*(m + n)).*** Although still a significant growth, ***our algorithm does not trigger an exponential growth of complexity that linear programming implies as the games grow larger.***

Hence the EPA approach has a lower computational complexity than linear programming for computing the Nash Equilibrium value of a two-person, zero sum game, particularly large games, as its computation of the column strategy is the simplest possible one, requiring no algebraic optimization that LP needs.

## 6.    Entropy

Entropy measures the uncertainty or randomness of a distribution and has been discussed by Cover, T.M., & Thomas, J.A. (2006). Playing each strategy with equal probability maximises the entropy of the defender's strategy. By maximising entropy, the defender makes it most difficult for the row player to predict the defender's strategy. We give mathematical proof that EPA maximises entropy while the entropy achieved by PSA is less than that achieved by EPA.

In this context, it is necessary to state the Gibbs inequality, also known as the Gibbs entropy inequality, a fundamental concept in information theory and statistics. Gibbs inequality states that: the entropy of a probability distribution is maximized when the distribution is uniform.

Mathematically, the Gibbs inequality can be expressed as:

$$H(P) \leq H(Q)$$

where:

- H(P) is the entropy of the probability distribution P
- H(Q) is the entropy of the uniform probability distribution Q
- P and Q are probability distributions over the same set of outcomes

**Proposition 4:** *The entropy of the PSA approach is less than the entropy of the EPA approach and equals that of EPA only when PSA is itself the uniform distribution. The entropy of the linear programming approach is also less than EPA unless the LP approach imposes a uniform distribution on column's strategy probabilities.*

**Proof:**

**Entropy of the Column Player's (defender's) Strategy:**

Given the column player's strategy as a probability distribution over their 'n' active strategies, $\sigma = (\sigma_1, \sigma_2, \dots, \sigma_n)$, where $\sigma_i$ is the probability of playing the $i$-th strategy, entropy $H(\sigma)$ is defined as:

$$H(\sigma) = -\sum_{i=1}^{n} \sigma_i \log_2(\sigma_i)$$

In the EPA, each of the 'n' active strategies is played with an equal probability. Therefore, for the EPA strategy, denoted as $\sigma_{EPA}$, the probability of each strategy $i$ is :

$$\sigma_{EPA,i} = \frac{1}{n} \text{ for all } i = 1, 2, \dots, n$$

The entropy of this EPA strategy is given by:

$$H(\sigma_{EPA}) = -\sum_{i=1}^{n} \sigma_{EPA,i} \log_2(\sigma_{EPA,i})$$

Substituting $\sigma_{EPA,i} = \frac{1}{n}$ :

$$H(\sigma_{EPA}) = -\sum_{i=1}^{n} \frac{1}{n} \log_2\left(\frac{1}{n}\right)$$

Since $\log_2\left(\frac{1}{n}\right) = \log_2(n^{-1}) = -\log_2(n)$, we rewrite as:

$$H(\sigma_{EPA}) = -\sum_{i=1}^{n} \frac{1}{n}(-\log_2(n))$$

$$H(\sigma_{EPA}) = \sum_{i=1}^{n} \frac{1}{n} \log_2(n)$$

Since $\log_2(n)$ is a constant with respect to the summation index $i$, we take it out of the sum:

$$H(\sigma_{EPA}) = \log_2(n) \sum_{i=1}^{n} \frac{1}{n}$$

Therefore, the entropy of the column player's strategy under the equal probability approach is:

$$H(\sigma_{EPA}) = \log_2(n)$$

**Entropy of the PSA Approach**;

Under the PSA approach, the column player plays each strategy with a probability that sums up to 1:

$$\sigma_{PSA} = (\sigma_1, \sigma_2, \ldots, \sigma_n)$$

where $\sum_{i=1}^{n} \sigma_i = 1$

The entropy of the PSA approach is:

$$H(\sigma_{PSA}) = -\sum_{i=1}^{n} \sigma_i \log_2(\sigma_i)$$

**The Gibbs inequality:**

For any two probability distributions $p = (p_1, p_2, \ldots, p_n)$ and $q = (q_1, q_2, \ldots, q_n)$ over the same set of events, the following inequality holds:

$$-\sum_{i=1}^{n} p_i \log_2(p_i) \leq -\sum_{i=1}^{n} p_i \log_2(q_i)$$

Equality holds if and only if $p_i = q_i$ for all $i$:

**Applying the Gibbs inequality to PSA and EPA:**

Setting $p = \sigma_{PSA}$ (any valid probability distribution) and $q = \sigma_{EPA}$ (the uniform distribution), applying the Gibbs inequality, we get:

$$-\sum_{i=1}^{n} \sigma_i \log_2(\sigma_i) \leq -\sum_{i=1}^{n} \sigma_i \log_2\left(\frac{1}{n}\right)$$

The left side of inequality is the entropy of the PSA strategy, $H(\sigma_{PSA})$. Simplifying the right side:

$$-\sum_{i=1}^{n} \sigma_i \log_2\left(\frac{1}{n}\right) = -\sum_{i=1}^{n} \sigma_i \left(-\log_2(n)\right)$$

$$= \sum_{i=1}^{n} \sigma_i \log_2(n)$$

Since $\log_2(n)$ is a constant with respect to the summation index $i$, we take it out of the sum:

$$= \log_2(n) \sum_{i=1}^{n} \sigma_i$$

Since $\sigma_i = \frac{1}{n}$,

$$\sum_{i=1}^{n} \sigma_i = 1$$

Therefore, the right side simplifies to:

$$\log_2(n)$$

So, the Gibbs inequality applied to PSA and EPA gives us:

$$H(\sigma_{PSA}) = -\sum_{i=1}^{n} \sigma_i \log_2(\sigma_i) \leq \log_2(n)$$

Since, $\log_2(n)$ is the entropy of the EPA strategy, $H(\sigma_{EPA})$, we can write:

$$H(\sigma_{PSA}) \leq H(\sigma_{EPA})$$

or equivalently:

$$H(\sigma_{PSA}) - H(\sigma_{EPA}) \leq 0$$

This shows that the entropy of the PSA approach is less than or equal to the entropy of the EPA approach. The entropy of PSA being equal to that of EPA would a limiting case where the random distribution of probabilities across the defender's active strategies fortuitously are all equal to one another.

The Gibbs inequality states that the entropy of a probability distribution is maximized when the distribution is uniform.

Since the EPA approach corresponds to a uniform distribution, its entropy is maximized consistently and not fortuitously as in the case of PSA. Maximum entropy is a structural property of EPA whereas it is not a structural property of PSA. Non-uniform distributions reduce entropy. Since PSA may assign unequal probabilities to the defender's active strategies with a far greater probability than equal probabilities, the entropy of PSA would decrease.

Therefore, we can conclude that:

The EPA approach maximizes the entropy of the defender's strategy set since the EPA approach corresponds to assignment of the defender's strategies over a uniform probability distribution.

In the context of the Colonel Blotto game as well as of all two person, zero sum games, the Gibbs inequality is used to show that the EPA approach (equal probabilities) maximizes the entropy of the column player's (defender's) strategy, making it the most uncertain or random. The above theorem also proves that the entropy of the Nash Equilibrium strategy of the defender computed by the Lemke-Howson algorithm or by a Linear Programming solver may be less than that of the equal probabilities strategy.

Since the EPA approach introduces more uncertainty and randomness than the PSA, the attacker would have a harder time identifying and exploiting any potential deviations from his Nash Equilibrium strategy to gain an advantage. The attacker would more likely stick with the Nash Equilibrium strategy, which results in the same game value. On the other hand, as PSA provides less uncertainty and therefore more predictability, it becomes easier for the attacker to identify and exploit deviations from his Nash Equilibrium strategy that lead to a potentially worse outcome for the defender.

The same argument would apply against a Linear Programming approach. LP solvers being deterministic, do not account for uncertainty or randomness in player behaviour. Hence they may not be able to mitigate the effects of Selten's Trembling Hand Theorem, which our algorithm can handle by having the column strategies played with equal probability. This is explained below. As the entropy of column's strategy distribution in LP based solutions is less than in our algorithm, row can exploit patterns in column's behaviour.

## 7.    Robustness to Deviation due to mistake:

Selten's Trembling Hand theorem states that if a Nash Equilibrium is not trembling hand perfect, then there exists a small probability of deviation from the equilibrium that can lead to a better payoff. This theorem highlights the importance of robustness to deviations.

In the context of the Colonel Blotto Game and two-person, zero sum games in general, we now prove that EPA provides column (defender) greater robustness of his payoff to deviations by the row player (attacker) from row's Nash Equilibrium strategies than PSA.

**Proposition 5:** *The equal probabilities approach (EPA) is more robust to the row (attacker) player's trembling hands deviations than the probabilities summing to 1 approach (PSA) or the Linear Programming approach.*

**Proof:**

We denote the column (defender) player's strategies as $C_1, C_2, \ldots, C_n$ and the row (attacker) player's strategies as $R_1, R_2, \ldots, R_m$. The payoffs for the defender are given by the matrix $P$, where $P_{ij}$ is the payoff for the defender when the defender plays $C_i$ and the row player plays $R_j$.

Suppose the row player's Nash equilibrium strategy is $\sigma_R^* = (\sigma_{R1}^*, \sigma_{R2}^*, \ldots, \sigma_{Rm}^*)$. According to Selten's trembling hands theory, the row player may deviate from his Nash equilibrium strategy with a small probability $\epsilon$. We denote the row player's trembling hands strategy as $\sigma_R^{TH} = (\sigma_{R1}^{TH}, \sigma_{R2}^{TH}, \ldots, \sigma_{Rm}^{TH})$, where:

$$\sigma_{Rj}^{TH} = \sigma_{Rj}^* + \epsilon_j$$

for some small $\epsilon_j$.

Now, we consider the defender's expected payoff under the equal probabilities approach:

$$E[P]_{EP} = \sum_{i=1}^{n} \frac{1}{n} \sum_j P_{ij} \sigma_{Rj}^{TH}$$

Where the term $P_{ij}\sigma_{Rj}^{TH}$ represents the expected contribution to the defender's payoff from the attacker playing strategy $R_j^{TH}$

Substituting, we get:

$$E[P]_{EP} = \sum_{i=1}^{n} \frac{1}{n} \sum_{j} P_{ij}(\sigma_{Rj}^* + \epsilon_j)$$

Using the linearity of expectation, we rewrite this as:

$$E[P]_{EP} = \sum_{i=1}^{n} \frac{1}{n} \sum_{j} P_{ij}\sigma_{Rj}^* + \sum_{i=1}^{n} \frac{1}{n} \sum_{j} P_{ij}\epsilon_j$$

The first term is the defender's expected payoff under the row player's Nash equilibrium strategy, which is a constant. The second term represents the effect of the row player's trembling hands deviation.

Now, we consider the probabilities summing to 1 approach:

$$E[P]_{PS} = \sum_{i} \sigma_{PS,i} \sum_{j} P_{ij}\sigma_{Rj}^{TH}$$

By the linearity of expectation, this may be rewritten as:

$$E[P]_{PS} = \sum_{i} \sigma_{PS,i} \sum_{j} P_{ij}\sigma_{Rj}^* + \sum_{i} \sigma_{PS,i} \sum_{j} P_{ij}\epsilon_j$$

The key differences lies in the $\frac{1}{n}$ term. This term implies that as the number of defender's strategies ($n$) increases, positive and negative errors tend to cancel out among an increasing number of $n$ strategies. On the other hand, use of a specific probability distribution over defender's strategies (the probability "summing to 1" approach) does not guarantee that the impact of the attacker's (row's) trembling hands would decrease when the defender has more strategies. The effect could be larger depending on how the probabilities $\sigma_{PS,i}$ are distributed and the specific values in the payoff matrix $P$. The probability distribution under the PSA approach may actually spike the effect of row's trembling hands; or alternately, completely neutralize it.

Therefore, we can conclude that the equal probabilities approach is more robust to the row player's trembling hands deviation than the probabilities summing to 1 approach.

Linear programming solvers suffer from the same sensitivity to trembling hand deviations as the PSA since they are deterministic and do not account for such trembles. They may require additional optimization techniques to handle such deviations to which they are not inherently robust.

## 8.     Minimax regret:

Regret is a measure of difference between the actual payoffs and the optimal payoff (Osborne, M. J., & Rubinstein. A, 1994. We give proof that playing each strategy with equal probability (EPA) has a simple closed form expression for the expected regret. We show, on the other hand, that the probabilities summing to 1 approach (PSA) does not have a simple closed form expression for the expected regret.

**Proposition 6:** *The equal probabilities approach (EPA) has a simple closed form expression for the expected regret while the probabilities summing to 1 approach (PSA) does not have. LP solvers too cannot handle non-linear regret functions.*

**Proof:**

Regret Minimization

Let's denote the column player's (defender) strategies as $C_1, C_2, \ldots, C_n$ and the row (attacker) player's strategies as $R_1, R_2, \ldots, R_m$. The payoffs for the column player are given by the matrix $P$, where $P_{ij}$ is the payoff for the column player when they play $C_i$ and the row player plays $R_j$.

The regret of the column player for playing strategy $C_i$ when the row player plays $R_j$ is given by:

$$Regret_{ij} = \max_k P_{kj} - P_{ij}$$

The expected regret of the column player for playing a mixed strategy σ is given by:

$$E[Regret] = \sum_i \sigma_i \sum_j P_{ij} Regret_{ij}$$

The expected regret is thus arrived at by weighing the regret incurred at each strategy $i$ by how likely was column to play $i$; and, inside that, by weighting columns regret across possible rows strategies $j$ by the column's actual payoff $P_{ij}$.

**Equal Probabilities Approach**

Let's denote the equal probabilities approach as $\sigma_{EP}, where\ \sigma_{EP,i} = 1/n$ for all $i$. Then:

$$E[Regret]_{EP} = \sum_i \frac{1}{n} \sum_j P_{ij} Regret_{ij}$$

Since the regret is a linear function of the payoffs, we can rewrite the expected regret as:

$$E[Regret]_{EP} = \frac{1}{n} \sum_j \sum_i P_{ij} Regret_{ij}$$

Now, we consider the optimal mixed strategy σ* that minimizes the expected regret. By the minimax theorem, we know that:

$$\min_{\sigma} \max_j \sum_i \sigma_i P_{ij} Regret_{ij} = \max_j \min_{\sigma} \sum_i \sigma_i P_{ij} Regret_{ij}$$

where,

$\sum_i \sigma_i P_{ij} Regret$ is the expected regret defender (column) suffers when row plays pure strategy $R_j$

$\max_j$ is row's worst case for column

$\min_{\sigma}$ is column's best mixed strategy to minimize this worst case expected regret inflicted by row.

Since the equal probabilities approach $\sigma_{EP}$ is a feasible solution, we know that:

$$\min_{\sigma} \max_j \sum_i \sigma_i P_{ij} Regret_{ij} \leq \max_j \sum_i \frac{1}{n} P_{ij} Regret_{ij}$$

The minimum overall strategies σ must be less than or equal to the value one gets from any one strategy, which is $\sigma_{EP}$ in this case. A symmetric or well-behaved game may even make $\sigma_{EP}$ near optimal.

Therefore, we can conclude that the equal probabilities approach has a simple closed form expression for the expected regret. It gives us a simple, closed-form upper bound on the expected regret when row plays the worst case for column.

**Probabilities Summing to 1 Approach**

We denote the probabilities summing to 1 approach as $\sigma_{PS}$, where $\sum_i \sigma_{PS,i} = 1$. Then:

$$E[Regret]_{PS} = \sum_i \sigma_{PS,i} \sum_j P_{ij} Regret_{ij}$$

Unlike the equal probabilities approach, the probabilities summing to 1 approach does not have a simple closed-form expression for the expected regret. A closed form expression is a mathematical expression that can be evaluated exactly using a finite number of operations, without requiring numerical approximations or iterative methods. EPA provides a closed form expression for the expected regret:

$$E[Regret]_{EP} = \frac{1}{n} \sum_{j=1}^{m} \sum_{i=1}^{n} P_{ij} Regret_{ij}$$

On the other hand, the PSA approach does not provide a closed form expression for the expected regret. It involves unknown weights $\sigma_{PS,i}$ that must be optimised to minimize regret. This makes the entire expression non-linear. The expected regret under the PSA approach would typically require approximations or iterative methods to evaluate, such as Monte Carlo simulations or iterative optimization algorithms. The PSA approach would involve optimizing over a set of probabilities that sum to 1, which can lead to complex optimization problems that do not have closed form solutions.

LP solvers, being designed to solve linear programming problems may not necessarily be designed to be applicable to minimax regret problems. Moreover, minimax regret often involves non-linear regret functions, which can be challenging for LP solvers to handle.

## 9.     Robustness to Incomplete Information

R.J. Aumann & M. Maschler (1995) have discussed the robustness of strategies to incomplete information. Playing each strategy with equal probability provides greater robustness to incomplete information. In the Colonel Blotto Game and two-person, zero sum games in general, the column player may not have complete information about the row player's strategy. By playing each strategy with equal probability, the defender can ensure that his strategy is robust to incomplete information. We now prove that EPA is more robust to incomplete information than PSA.

**Proposition 7:** *EPA is more robust to incomplete information than PSA or Linear Programming (which assumes complete information about problem parameters).*

**Proof**

Definitions:

Let $A = \{C_1, C_2, \ldots C_k\}$ be the active support of column's strategy.

**EPA strategy:**

$$\sigma^{EPA} = \left(\frac{1}{k}, \ldots, \frac{1}{k}\right) \ over \ A$$

**PSA Strategy:**

$$\sigma^{PSA} = \left(\sigma_{1,\ldots}\sigma_K\right) over \ A, \ \sum_{i=1}^{k} \sigma_i \ = 1, \sigma_i \geq 0$$

Now, let column be uncertain about which Row strategy is being played. Let: $\tau_1, \ldots \tau_m$ be possible mixed strategies used by Row

Each $\tau_j \ \in \ \Delta_r$, where $\Delta_r$ is the simplex over Row's pure strategies

Row plays $\tau_j$ with probability $p_{j,}$ where $\sum_{j=1}^{m} p_j \ = 1$ .

Let the payoff matrix between row's pure strategies and column's active pure strategies be:

$M \in R^{r \times k}$ (rows = Rows pure strategies, columns = Column's active strategies)

Then the expected payoff vector to column, when row plays $\tau_j$, is :

$P^{(j)} = \ \tau_j^T \ M \ \in R^k$ where $P_i^{(j)}$ is the expected payoff for column's strategy $C_i$

Let column play strategy $\sigma \in \ \Delta_k$. The expected payoff if row plays $\tau_j$, is:

$$u_j \ (\sigma) = \sum_{i=1}^{k} \sigma_i \ P_i^{(j)}$$

But the best column could have done (had column known $\tau_j$) is:

$$\max_i P_i^{(j)}$$

So the average regret under uncertainty caused by incomplete information available to column (defender) about row's probability distribution is:

$$R(\sigma) = \sum_{j=1}^{m} p_j \left[ \max_i P_i^{(j)} - \sum_{i=1}^{k} \sigma_i P_i^{(j)} \right]$$

Where,

$\max_i P_i^{(j)}$ is the best column could have done against $\tau_j$;

$\sum_{i=1}^{k} \sigma_i P_i^{(j)}$ is what column actually gets by playing mixed strategy

The difference is column's regret for not knowing $\tau_j$

$\sum_{j=1}^{m} P_j$ averages this regret over the unknown distribution of row's mixed strategy profiles (sets), $m$

Since the first $\max_i P_i^{(j)}$ term is constant, minimizing regret means maximizing:

$$\sum_i^{k} \sigma_i P_i^{(j)}$$

And this is maximized when $\sigma_i = \frac{1}{k}$, i.e. under EPA. Any other skew (PSA) might reduce the expected value of the game by exacerbating the effect of the deviation of the presumed Nash Equilibrium probabilities of row from the actual row NE probabilities due to column's ignorance. EPA is the safest method to adopt by the defender when he has incomplete information about the attacker's exact strength, which influences payoffs and thus the computation of the row NE probabilities, – a realistic scenario not only in the fog of war but also at normal times.

LP solvers assume complete information about problem parameters, which is obviously not a realistic assumption in many situations. Hence LP solvers would probably not be able to find the optimum solution to problems with incomplete information, leading to sub-optimal decisions.

It will be seen from the foregoing pages that the applicability of the proposed algorithm is not restricted to only two-person, zero-sum Colonel Blotto games, which is, in its traditional form, a special kind of the two-person, zero-sum game in general. Instead, the algorithm can be applied to all two-person, zero-sum games where the Nash equilibrium is typically and conventionally computed by spreading the active strategy set of each player over a Nash-determined probability distribution.

## 10. Discussion

This asymmetric equilibrium idea where row plays Nash while column merely neutralizes by uniformizing over the strategies row made indifferent is novel and unprecedented with no canonical reference (eg. Myerson, Hart, Osborne and Rubenstein). Our proposed method resembles a reactive defence, not a strategic counter-optimization. It bypasses the usual requirements that both players solve linear programmes to play Nash. Our method is not a standard dual formation, not is it a best-response in the conventional sense. Yet it has strategic stability.

It yields the same game value for column with less computational effort. (It requires the least computational effort compared to any other extant method)

It drastically reduces effort on the part of column (defender) which is not required to compute a full NE. In the standard two-person, zero sum game, (including Colonel Blotto) the value of the game is typically guaranteed only when both players play NE strategies. (Roberson, Hart etc.) *On the other hand, if column plays each of his active defensive strategies with an equal probability, row would be forced to play Nash to optimize his outcome.*

Column's behaviour resembles a "satisficing" or bounded rationality. It's a provable strategy, yet it respects the asymmetry of strategic depth that is common in real-world attacker-defender situations.

We have given a clear mathematical proof that shows: (i) the neutralizing column strategy caps row's expected payoff at the same value row would get in Nash play. (ii) column loses nothing by not computing a full NE.

| No. | Feature | Why it is valuable |
|-----|---------|--------------------|
| 1. | Asymmetric modelling | Breaks standard assumptions, introducing new insight |
| 2. | Lower complexity | Operationally optimal in defense/ resource allocation |
| 3. | Formal guarantee (proof) | Provides rigour beyond heuristic or ad-hoc models |
| 4. | Not covered in existing literature | Original contribution to two-person, zero sum games & Colonel Blotto games |
| 5. | Real world alignment | Reflects bounded rationality in human/ organizational defenders |

Such an equilibrium may be christened a '**Reactive Defense Equilibrium**' (**RDE**).

In the fast-evolving landscape of conflict, both conventional and irregular, strategic resource allocation for defense using the discrete Colonel Blotto game framework is of special importance: it assures the defender of a reasonably certain payoff in terms of the value of the game assuming his reasonable (even if not strictly accurate) estimate of the attacker's strength. Despite the claim that linear programs can be solved efficiently, this author has empirically ascertained that LP solvers do not solve large Colonel Blotto games in polynomially efficient time. Compared to the Lemke Howson algorithm, which uses the Pivoting Path method, or LP solvers that algebraically optimize the strategies of both row and column, the proposed algorithm offers a simple alternative to the NE computation by algebraic optimization of column (defender) strategies, apart from the other considerably important advantages described. With less time to computation of the values of the game as a result of this simplification compared to the performance of LP solvers and Lemke Howson, the playing of the defender's un-dominated strategies (resource allocations across any given number of battlefields) with an equal probability would obviously be easier to implement than a catalogue of awkward, uneven probabilities.

Thus our proposed approach is peculiarly suited to the bounded rationality conditions of current warfare, where defensive Surface to Air Missile batteries, Directed-Energy Weapons and other Air Defense Systems will be required to be allocated across protected cities against airborne threats like incoming missiles, loitering munitions and a broad spectrum of Unmanned Aerial Systems. In irregular warfare and counterinsurgency, the proposed approach would equally well fit into the defender counterinsurgent's allocation of his resource units (troops, munitions, etc) across his protected locations to optimize his payoff against guerrilla attacks.

## 11. Future Research

The reduced computational complexity of the proposed algorithm, which computes the values of the Nash equilibrium in two-person, zero-sum games, including the Colonel Blotto game, has been argued in the paper. Its implementation would enable a precise computation of its performance metrics and a comparison with the performance achieved by linear programming solvers and other extant programming implementations of large zero sum, two-person games that compute the Nash Equilibrium strategy of both players. On the theoretical side, since the proposed algorithm cannot be characterized by the Fixed Point Theorems which have been shown to be in PPAD, further investigation is warranted to prove or disprove that the proposed algorithm is in PPAD. The proposed algorithm can also hold significance for mechanism design of second price auctions which can be described in terms of *n* person Colonel Blotto games.

# Appendix 'A'

**Proof of proposition 1:**

Let:

A = {a$_1$, a$_{2,....}$ a$_n$}: Row's pure strategies

B = {b$_1$, b$_2$,.... b$_n$}: Column's pure strategies

M ∈ $R^{m \times n}$: Row's payoff matrix. Column's payoff is − M; zero sum game.

Let:
$\Delta (A)$ : the simplex of probability distributions over $A$

$\Delta (B)$ : the simplex of probability distributions over $B$

## Definitions

Definition 1: (Row's Nash strategy over column's set of active strategies B$^*$)

Let $P^* \in \Delta (A)$ be a mixed strategy set such that there is a subset $B^* \subseteq$ B, with $|B^*| = K$, for which :

$$\forall b_j \in B^*, \ E_{P^*} [M:j] = V, \qquad \text{for some constant V,}$$

and for all $b_j \notin B^*$,

$$E_{P^*} [M:j] \leq V$$

This ensures that column is indifferent over all $b_j \in B^*$ if row plays $P^*$.

Definition 2:

Let column play the uniform strategy over the indifferent support:

$$q^*(j) = \begin{cases} \frac{1}{k}, & \text{if } b_j \in B^* \\ 0, & \text{otherwise} \end{cases}$$

Column does not attempt to make row indifferent across A. Instead, he neutralizes row by not deviating from the equal-weight support B$^*$.

## **Proposition:** (Neutralization guarantee)

Let $P^* \Delta (A)$ be a strategy set for row to make column indifferent across B$^*$, and let q$^* \in \Delta$ (B) be column's uniform distribution over B$^*$. Then

$$\forall p \in \Delta(A),\ E_{p,q*}[M] \leq E_{p*,q*}[M] = V$$

that is, column's neutralizing strategy q* minimizes row's expected payoff against row's own Nash strategy P*, and no deviation P≠ $P*$ by row can do better.

We prove this in two parts:

(i)     Expected payoff under P* and q*

By assumption, for all $b_j \in B*$, we have:

$$E_{P*}[M:j] = V \ldots \ldots \text{eq 1}$$

And

$$E_{p*,q*}[M] = \sum_{j \in B*} q*(j).E_{P*}[M:j] = \sum_{j \in B*} \frac{1}{K}.(V) = V \ldots \ldots \text{eq 2}$$

The same results of equations 1 and 2 (i.e. V) proves my proposition that:

$$u_i(a*) = u_i(a_i, a*_{-i})\ \text{for every action}\ a_i\ \text{of player i}$$

where,

Column player $C$ (defender) represent player $i$;  $a_i$ is a set of active strategies $n$ in a mixed strategy game such that each element of the set of active strategies $a_i = \{x_1, x_2 \ldots x_n\}$ is played with the probability 1/n; and

$a*$ is the action profile of each player given by the Nash equilibrium wherein the active strategies of each player are spread over a probability distribution *determined by the Nash equilibrium.*

Expected payoff under arbitrary P ∈ Δ (A)

We now show that for any P≠ $P*$, the expected payoff against q* is at most V.

Let     $V_j(P) = E_P[M:j]$, the expected payoff to Row if Column plays pure strategy $b_j$ . Then:

$$E_{p,q*}[M] = \sum_{j \in B*} \frac{1}{K}.V_j(P)$$

Recall:

(i) $V_j(P*) = V$ for all $j \in B*$

(ii) By construction of Nash strategy $P*$, row is guaranteed the payoff of at least V no matter what column plays within the support $B*$ (i.e. against any response over $B*$)

Now,

$$f(P) := \sum_{j \in B^*} \frac{1}{K} . E_P[M:j] = E_{p,q*}[M]$$

That is, this is the expected payoff to row when he plays P and column plays q*, the uniform distribution over B*.

Since, P* is a Nash strategy, it is optimal for row, guaranteeing value V against any column strategy supported on B*.

In particular, it maximizes $f(P)$

So:

$$f(P) \leq f(P^*) = V \text{ for any } P \in \Delta(A)$$

Therefore, by the convexity of expression,

$$\sum_{j \in B^*} \frac{1}{K} . V_j(P) \leq \sum_{j \in B^*} \frac{1}{K} . V = V$$

That is,

$$\sum_{j \in B^*} \frac{1}{K} . V_j(P) = E_{p,q*}[M] \leq E_{p*,q*}[M] = V$$

Hence,

$$E_{p,q*}[M] \leq E_{p*,q*}[M]$$

## Deviation of the proposed algorithm from the framework of Tarski, Kakutani and Brouwer Fixed Point Theorems

It is first proposed to show that the Tarski Fixed Point Theorem cannot be applied to the proposed algorithm despite the latter computing the same value of the two person zero sum Colonel Blotto game as the Nash Equilibrium solution. It will then be shown that the proposed algorithm does not fit into the framework of the Kakutani Fixed Point Theorem either, which like the Tarski Fixed Point Theorem is applied to discrete strategy spaces (like the original Colonel Blotto Game). Finally, by extending the two-person Colonel Blotto Game to continuous strategy spaces it would still be shown that the proposed algorithm deviates from the framework of the Brouwer Fixed Point Theorem as well.

**Proposition:** *The proposed algorithm computes the same value of a two-person zero sum Colonel blotto game (and of any other two-person zero sum game), as the Nash Equilibrium solution in spite of deviating from the framework and conditions of the Tarski Fixed Point Theorem.*

**Proof:**

Tarski's theorem states that if we have a monotone function (or correspondence) on a complete lattice, then there is a Fixed Point, $\sigma = f(\sigma)$. In the context of discrete strategy spaces, a complete lattice represents the discrete strategy space and a monotone function represents the best-response correspondence between players. The three requirements of the theorem are therefore: (a) Complete Lattice; (b) Monotone Function $[x \leq y \implies f(x) \leq f(y)]$; & (c) Order-Preserving.

In the context of our Police-Guerilla Colonel Blotto Game (and by implication, for every two-person zero sum game), to apply Tarski's Fixed Point Theorem, it is necessary to form a complete lattice. To form a complete lattice, every subset of the strategy space must have (i) a **least upper bound** and a **greatest lower bound**; and (ii) the ordering of the strategies of both players must be consistent with the pay-offs.

This means that the combined strategy space S = L x G (where L represents the guerilla's strategies and G represents the Police' strategies) does not satisfy the complete lattice requirements if:

i)      The least upper bound (**lub**) and greatest lower bound (**glb**) properties do not hold for subsets of S, OR

ii)     The ordering of the strategies of both players is not consistent with the pay-offs

In the present case, column's (Police') strategy subset is *not closed* under the Least Upper Bound and Greatest Lower Bound; since column is now constrained to equal probabilities strategies on a subset, that subset breaks closure under lub/glb. This violates the complete lattice requirement.

The combination of a Nash-determined (possibly uneven) probability distribution for the Guerilla's (attacker's) strategies and a uniform (non-Nash) probability distribution for the Police' (defender's) strategies leads to a situation where (i) the guerilla's (attacker) strategies is ordered consistently with the pay-offs; (ii) The Police' (defender's) strategies is not necessarily ordered consistently with the pre-equalization pay-offs.

This inconsistency also violates the complete lattice requirements making it impossible to apply Tarski's Fixed-Point Theorem.

To define monotonicity, a partial order on column (defender's) strategies is needed. However, in Police' (defender's) strategy space of equal probabilities there is no strategic ordering that is compatible with the pre-equalization payoff outcomes of the defender. A change of probabilities in the column active strategy set from equal probability values (required by our approach) to Nash Equilibrium values leaves row's best response (Nash) unchanged, which violates the monotonicity requirement of Tarski.

Mathematically,

Let:

$\tilde{\sigma}_c \in \Delta_n$ be column's uniform strategy over active pure strategies $S \subseteq \{1,......,n\}$

$\sigma_c^* \in \Delta_n$ be column's Nash strategy (on the same support S) with NE weights

$\sigma_R^* \in \Delta_m$ is row's best response to both strategies

Let f: $\Delta_n \to \Delta_m$ be row's best response map.

Now, **Input variation:** $\tilde{\sigma}_c \neq \sigma_c^*$ but supp $(\tilde{\sigma}_c) = $ supp $(\sigma_c^*)$

The two strategies differ in weights but share the same set of active pure strategies. They lie on the same face of the simplex $\Delta_n$, but with different interior points.

**Constant best response:**

$$f(\tilde{\sigma}_c) = f(\sigma_c^*) = \sigma_R^*$$

Despite different inputs, Row responds with the same mixed strategies – no change in the best response.

This makes f constant on that face of the simplex.

**Monotonicity failure:**

Suppose $\preceq$ is any partial order on $\Delta_n$ intended to support Tarski's theorem. Then Tarski requires:

$$\tilde{\sigma}_c \neq \sigma_c^* \implies f(\tilde{\sigma}_c) \neq f(\sigma_c^*)$$

or at least:

$$\tilde{\sigma}_c \preceq \sigma_c^* \implies f(\tilde{\sigma}_c) \preceq f(\sigma_c^*)$$

But we have:

$$\tilde{\sigma}_c \neq \sigma_c^*$$

$$\text{yet } f(\tilde{\sigma}_c) = f(\sigma_c^*)$$

So no matter how $\tilde{\sigma}_c$ and $\sigma_c^*$ is ordered, Row's response is not sensitive to that ordering. Hence, f is not monotonic on this subset. Therefore, Tarski fails.

Thus, our proposed algorithm cannot be deterministically characterized by the Tarski Fixed Point Theorem.

**Corollary**: As Fixed-Point Theorems like Tarski, Kakutani & Brouwer have been reduced to END OF THE LINE in PPAD, it follows that our proposed algorithm cannot be held to be in PPAD solely on the basis of the reduction of TARSKI to END OF THE LINE in PPAD. (In 2008, Etessami & Yannokakis showed that the Tarski Fixed Point Theorem could be reduced to the END OF THE LINE argument in PPAD).

**Proposition**: *The proposed algorithm computes the same value of a two-person, zero-sum Colonel Blotto Game (and of any other two-person zero sum game) as the Nash equilibrium solution in spite of deviating from the framework and conditions of the Kakutani Fixed Point Theorem.*

**Proof:**

Our algorithm involves:

i)   The Police (defender) strategies are played over a uniform (not Nash) probability distribution.

ii)  The Guerilla (attacker) strategies are played over a (possibly uneven) Nash determined probability distribution.

Kakutani's Fixed Point Theorem requires:

1. A compact, convex set.
2. A non-empty, convex-valued correspondence.
3. A closed graph.

If, column (defender) is allowed to play Nash Equilibrium strategies, column's strategy space becomes the full probability simplex:

$$\Delta^n = \{ y \in R^n \mid y_i \geq 0 \sum_{i=1}^n y_i = 1 \}$$

This space contains infinitely many mixed strategies. Taking any two mixed strategies $y^{(1)}$, $y^{(2)} \in \Delta^n$, then for any $\boldsymbol{\lambda} \in [0, 1]$,

$$y = \boldsymbol{\lambda} \, y^{(1)} + (1 - \boldsymbol{\lambda}) \, y^{(2)} \in \Delta^n$$

Therefore, $\Delta^n$ is convex. It satisfies the Kakutani convexity condition.

When column (defender) is restricted to uniform distribution over subsets $S \subseteq C$ of fixed size k, so that each strategy in S is played with probability $1/k$, and all others with probability 0:

For each such S, there is only one allowed strategy:

$$y_S^* \text{ with } (y_S^*) i = \begin{cases} \frac{1}{k}, & i \in s \\ 0, & otherwise \end{cases}$$

So the full strategy space becomes:

$$U_k = \{ y_S^* \in \Delta^n \mid S \subseteq C, \mid S \mid S = k \}$$

This set has exactly $\binom{n}{k}$ isolated points - no interior.

Taking any two distinct elements $y_{s1}^* \; y_{s2}^* \in U_k$, their convex combination

$$z = y = \boldsymbol{\lambda} \, y_{s1}^* + (1 - \boldsymbol{\lambda}) \, y_{s2}^*$$

is not in $U_k$ because the support z is generally larger than k and the probabilities are no longer equal across support. This makes z a non-uniform distribution over any size - k subset. Thus $U_k$ is not convex. Our framework breaks convexity by design because even though $y_{s1}^*$ and $y_{s2}^*$ are allowed individually, their average z is not allowed.

From this, it follows that our proposed algorithm cannot be deterministically characterized by the Kakutani Fixed Point Theorem.

**Corollary:** As the Kakutani Fixed Point Theorem has been reduced to the END OF THE LINE argument in PPAD by Papadimitriou (1994), our proposed algorithm cannot be held to be in PPAD solely on the basis of the reduction of Kakutani to END OF THE LINE in PPAD.

**Proposition:** *The proposed algorithm computes the same value of a two person, zero-sum Colonel Blotto Game (and of any other two-person zero sum game) as the Nash equilibrium solution in spite of deviating from the framework and conditions of the Brouwer Fixed Point Theorem.*

**Proof:**

Brouwer's Fixed Point Theorem requires a continuous function from a convex, compact set to itself.

Kakutani's Fixed Point Theorem, on the other hand, allows for non-continuous (discrete) functions, but requires a convex-valued correspondence.

Although two person, zero-sum games like the Colonel Blotto game are typically played in a discrete strategy space, they can be easily extended to a continuous strategy space.

A Brouwer Graph must have a trichromatic triangle, which indicates a fixed point when:

1. The graph represents the strategy spaces of two players (in this case Guerilla & Police)
2. The players' strategies are INTERDEPENDENT, meaning that the optimal strategy for one player depends on the strategy chosen by the other player.

This would require, in the Guerilla-Police game:

1. The guerilla's strategy depends on the Police' strategy.
2. The Police strategy depends on the guerilla's strategy.

This interdependence creates a circular relationship between the 2 players' strategies.

When we represent this interdependence graphically, using a Brouwer graph:

1. Each point in the graph corresponds to a pair of strategies, one for the guerrilla and one for the police.
2. The colours of the graph (typically red, blue and green) represent the directions of the vector field, indicating how each player's strategy changes in response to the other player's strategy.

The trichromatic triangle appears when:

1. The interdependence between the players' strategies creates a "cycle" in the graph, where each player's strategy depends on the other player's strategy.
2. This cycle creates a "triangle" shape in the graph, where each vertex represents a pair of strategies.

The trichromatic triangle indicates a fixed point because:

1. The fixed point represents a Nash equilibrium, because both players' strategies are mutually best responses.
2. The trichromatic triangle has a vector field of '0' indicating a fixed point.

It follows that because in our algorithmic approach, there is no interdependence between the guerrilla's strategy and the Police strategy (as the guerrilla's active strategies are spread over a Nash determined, possibly uneven probability distribution while the Police' active strategies are spread over a uniform, non-Nash determined probability distribution), therefore our algorithm cannot be characterized by a Brouwer Graph with a fixed point i.e by the Brouwer Fixed Point Theorem.

**Corollary:** As the Brouwer Fixed Point Theorem has been reduced to the END OF THE LINE argument in PPAD by Daskalakis et.al (2008), our proposed algorithm cannot be held to be in PPAD solely on the basis of the reduction of BROUWER to END OF THE LINE in PPAD.